# Securing the Digital World: Protecting smart infrastructures and digital industries with Artificial Intelligence (AI)-enabled malware and intrusion detection

*Marc Schmitt*

*Department of Computer Science, University of Oxford, UK*

## Abstract

The last decades have been characterized by unprecedented technological advances, many of them powered by modern technologies such as Artificial Intelligence (AI) and Machine Learning (ML). The world has become more digitally connected than ever, but we face major challenges. One of the most significant is cybercrime, which has emerged as a global threat to governments, businesses, and civil societies. The pervasiveness of digital technologies combined with a constantly shifting technological foundation has created a complex and powerful playground for cybercriminals, which triggered a surge in demand for intelligent threat detection systems based on machine and deep learning. This paper investigates AI-based cyber threat detection to protect our modern digital ecosystems. The primary focus is on evaluating ML-based classifiers and ensembles for anomaly-based malware detection and network intrusion detection and how to integrate those models in the context of network security, mobile security, and IoT security. The discussion highlights the challenges when deploying and integrating AI-enabled cybersecurity solutions into existing enterprise systems and IT infrastructures, including options to overcome those challenges. Finally, the paper provides future research directions to further increase the security and resilience of our modern digital industries, infrastructures, and ecosystems.

# 1 Introduction

## 1.1 Securing the Digital World

The world is digital and complex. Our infrastructures and industries have become increasingly decentral, smart, and interconnected [1–5]. An ever-increasing basket of critical and emerging technologies will keep shaping the world we live in and continue to transform companies, countries, and entire societies [6]. We are close to industry 5.0 and smart devices have started to seamlessly bridge the physical and digital worlds [7].

This paper will focus on the security aspects of modern technologies such as communication networks [8], mobile devices [9], the internet of things (IoT) [10] , and cyber-physical systems (CPS) [2,11]. Those are the primary underpinnings and enablers of the digital economy and associated concepts such as smart infrastructures [12], smart cities [3], smart manufacturing [13,14], and the (industrial) metaverse [15]. Modern digital industries are heavily using industrial IoT ecosystems for digital twins, smart manufacturing, and digital supply chains [16–19]. The application areas for IoT are vast and offer significant economic value potential across an array of industries such as healthcare [20,21], agriculture [22], transportation, retail, energy management, product development, and security [23]. The dependence on those digital technologies continues to surge at a rapid rate. Unfortunately, the same is true for cybercrime. The modern hyperconnected world is the perfect playground for cybercriminals [24]. Cyber threats have grown into an omnipresent reality that endangers all stakeholders in our global economy. The world economic forum ranks cyber threats consistently among the global top security risks [25]. Cybercrime is growing rapidly in volume and magnitude causing severe monetary damage for corporates, violating the privacy of individuals, and has the potential to cause the breakdown of our critical infrastructures, or the disruption of vital supply chains by hacking smart factories.

Protecting our highly connected and technology-dependent world is vital for all stakeholders in our global economy. Especially small nations, SMEs, and individuals will have difficulties protecting themselves due to lacking resources and the missing, but necessary skillsets to build cyber resilience [26]. Cybersecurity experts are a rare commodity. Governments and large corporates are not secure either. Data and information are valuable and even insignificant vulnerabilities could potentially lead to huge losses for corporations and government entities due to malicious attacks or imprudent behavior, which could lead to confidentiality, integrity, or availability violations.

The digitalization of the world economy comes with increased security risks due to an ever-increasing attack surface. Cybercriminals use this constant innovation to actively search for new attack vectors. Cybersecurity today is best described as a never-ending arms race, where attackers and defenders are in a constant battle to develop new attacks and defenses, respectively.

So, how do we achieve cyber resilience? The answers are not simple, and a bullet-prove solution does not exist, but thinking about solutions and actively raising awareness of this important and interesting topic is vital for the security of industries, infrastructures, the world, and all its individuals.

## 1.2 AI in Cybersecurity

One of the core technologies of our time is artificial intelligence (AI) and machine learning (ML), which have found many use cases across industries, infrastructures, and cities [27,28]. AI/ML has also found several use cases in cyberspace and can help to shield systems from cyber-attacks in many ways [29]. Examples are network intrusion detection, malware detection, spam detection, network traffic analysis, etc. Concrete, applications of AI/ML in cybersecurity can mitigate emergent cyber threats in three ways: robustness, response, and resilience [30].

(1) Robustness: AI-based cyber security systems remain stable when faced with adversarial attacks (self-healing, self-testing).

(2) Response: AI-based cyber security systems are adaptive and able to learn from each attack to improve defense capabilities autonomously (e.g., launch counter operations, and generate decoys and honeypots).

(3) Resilience: AI helps cyber security systems to endure an attack. One of the key elements of resilient cyber security systems is the successful identification of cyber threats. This is done by AI-based cyber threat and anomaly detection systems/algorithms.

This paper focuses on analyzing and improving the resilience of AI-enabled anomaly-based intrusion detection systems, including the integration challenges into complex IT infrastructures.

### 1.2.1 Intrusion Detection Systems

The primary goal of Intrusion Detection Systems (IDS) is to automatically detect events that indicate attacks by malicious adversaries. As shown in Figure 1 IDS can be categorized into signature-based and anomaly-based methods. Anomaly-based methods focus on identifying deviations from the norm, while signature-based methods identify and match attack signatures [31]. The biggest problem with signature-based IDS is its inability to identify new attacks. Adversaries have created sophisticated malware that utilizes concealment techniques that makes systems alterations (malicious actions) often difficult to identify. For example, metamorphic malware attacks automatically reprogram themselves with each iteration. In contrast, anomaly-based detection algorithms based on AI/ML have a chance to identify malicious activities in all forms in a computer system. Higher detection accuracy increases the overall resilience of the security system and will be the primary focus of this paper. However, machine learning-based classification models are vulnerable to adversarial attacks [32]. Adversaries can target the ML-based intrusion detection system to mislead the classifier [33].

Another distinction can be made between network and host-based IDS. Network-based intrusion detection systems (NIDS) are largely concerned with inbound and outbound traffic monitoring and focus on network-related attacks. Host-based intrusion detection systems (HIDS) search for malicious traces and artifacts left behind by adversaries upon a successful cyberattack internally, but can also analyze the traffic on its network interfaces. One of the primary goals of an attacker is to keep the tunnel open by installing malware (e.g., backdoors, keyloggers, spyware) that might also turn the device into a zombie for a botnet and can then be used to carry out future attacks on other devices.

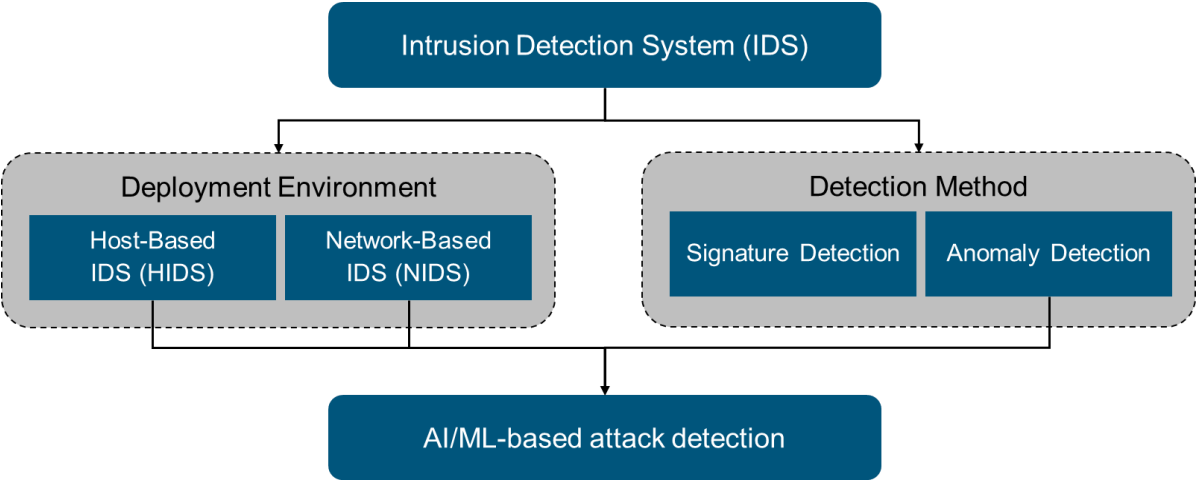

**Figure 1.** Modern Intrusion Detection Systems (IDSs) use AI/ML models for anomaly-based attack detection with the ability to identify also novel or zero-day threats. IDSs are passive security solutions, which identify anomalies and label them as potential threats for investigation by a security analyst. The accurate detection of threats is vital and the primary challenge of those models is to balance false positives with false negatives.

Both, network-based and host-based systems should work together. The NIDS analyses the traffic and tries to shield the system at the entrance. The HIDS will analyze the system and captures malicious activity that could not be prevented by the NIDS. The scenarios in this paper will cover both – network and host-based IDS.

### 1.2.2 AI-enabled Intrusion Detection

AI/ML-based anomaly and intrusion detection is an important and active field of research. Modern AI/ML-based intrusion detection systems offer robust protection against cyber-attacks by identifying anomalies and malicious behavior. The capability to detect zero-day attacks gives them an edge over traditional security methods [34]. A variety of machine learning techniques are applied in this area, including classical ML models like Naïve Bayes, decision trees, support vector machines, ensemble learning algorithms such as random forest and gradient boosting, and deep learning models with different architectures, such as MLP, CNN, and RNN [29]. However, there is no clear consensus on the best classifiers. While Abdullahi et al., [35] found support vector machines and random forests to be superior due to their high accuracy, Zhang et al., [36] found ensemble models performed well, while deep learning lagged. Contrarily, Fatani et al., [37] found deep learning to yield higher accuracy than other methods, and Imrana et al., [38] reported a LSTM-based system outperformed conventional LSTM and other state-of-the-art models.

A key focus area of current security research is the protection of industry 4.0 technologies like IoT and CPS [10,39,40]. CPSs are integrated technologies and find applications in many industries such as healthcare IoT, industrial IoT, smart cities IoT, etc. Studies by Medjek et al., [41] and Fatani et al., [37] showed mixed results for AI/ML-based intrusion detection in IoT environments. Decision trees, random forests, and K-Nearest Neighbors performed well, while deep learning, MLP, Naïve Bayes, and Logistic Regression showed lower performance. However, a single-layer neural network was found effective in a resource-constrained environment [42]. Fusion methods, such as stacking have shown to outperform base classifiers [43,44]. Model interpretability is another crucial research aspect. Rudin [45] argues against using black-box models for high-stakes decisions in many fields, however, there is a tradeoff between prediction accuracy and model explainability. For example, despite decision trees being less powerful in terms of prediction accuracy, they are valuable for their explainability and lower computational expense. The blackbox issue in the field of AI/ML is well-known, which has led to a surge in explainable AI (XAI) research in recent years.

Another important aspect of AI/ML models is their robustness against adversarial attacks. Studies by Tcydenova et al., [32] and Jmila and Khedher [33] indicate that a balance between resilience and robustness is necessary. Overall, the primary challenges of AI/ML-based intrusion detection systems are resilience (prediction accuracy), robustness (stability against adversarial attacks), and model explainability (blackbox). There is no superior AI/ML model that could be identified as a clear winner.

## 1.3 Digital Ecosystem Integration

Securing complex digital ecosystems is essential for our digital future and the integration of AI-enabled intrusion detection systems into existing IT/cybersecurity infrastructure is crucial for enhancing the detection and mitigation of cyber threats. However, integrating those models into complex enterprise systems presents several challenges. Advances in hardware, software, wireless communication, mobile applications, and IoT technologies have made it possible to incorporate IoT solutions into enterprise systems. IoT devices collect, transmit, and process data from various sources, providing valuable insights that can improve processes and decision-making. This increased data availability – from data science perspective – is good news for data-driven AI/ML models. However, ML applications are often considered "black boxes" due to their data-driven nature, making it difficult to understand their internal workings or predict their behavior. Overall, integrating AI/ML applications for cybersecurity presents unique challenges across the dimensions of process, data, application, and infrastructure. In addition, the presented primary challenges of ML applications, such as resilience, robustness, and explainability, and their overall data-driven nature, further exacerbate these integration complexities.

## 1.4 Research Questions and Scope

This research paper strikes directly at the heart of the world's technology needs and combines modern AI/ML as deep learning with cyber threat detection including the integration challenges in complex digital ecosystems. The need for intelligent systems to protect our heterogeneous digital industries and infrastructures with improvements in AI-based intrusion detection and prevention systems is vital to secure our future. Based on the literature review the following concrete research questions have been defined:

- RQ1: What is the best AI/ML model for anomaly-based supervised intrusion detection systems to detect cyber-attacks on mobile and IoT-enabled smart devices?
- RQ2: What are the primary challenges when integrating AI-enabled cybersecurity models in complex enterprise ecosystems?
- RQ3: How can we successfully integrate AI/ML-based cyber threat detection systems into complex digital ecosystems?

Digital infrastructures and ecosystems, including connected smart devices and communication networks, have complex interdependencies, bridge the physical and digital worlds, and connect legacy IT and modern technologies. This paper explores AI-based cyber threat detection to protect our modern digital ecosystems, we will focus on AI/ML applications in cybersecurity with a specific focus on network security, mobile security, and IoT security. Intrusion detection systems, whether for the network or the host system are the standard methods to shield entities from all kinds of cyber-attacks, however, integrating those systems into modern digital ecosystems poses several challenges. The following application scenarios will build the foundation of this paper:

- Scenario 1: Network Security - Network Intrusion Detection
- Scenario 2: Mobile Security - Android Malware Detection
- Scenario 3: IoT/CPS Security - IoT Cyber Threat Detection

Those three scenarios and datasets have been carefully selected to offer a balanced picture of cybersecurity for our modern digital industries and smart infrastructures. A vital aspect of AI/ML models for intrusion detection is detection accuracy and hence the reliability of the security solution. From an

ML perspective stacked ensembles will be used to increase threat detection accuracy by combining several base classifiers such as random forest, gradient boosting, and deep learning to a so-called super learner. Overall, the primary objective of this research is to develop an understanding of the integration challenges faced by AI/ML-enabled threat detection systems and contribute valuable insights into effective strategies and methodologies for addressing these challenges. Specifically, how can the integration challenges of ML-enabled cyber threat detection be effectively addressed to overcome potential obstacles in the process, data, application, and infrastructure dimensions? Industrial Information Integration Engineering (IIIE) as a relatively new discipline that focuses on how to integrate and manage various systems and data in an organization. Utilizing the components of IIIE could help address the challenges associated with integrating AI-enabled cybersecurity models into complex enterprise systems.

The structure of this paper is as follows: Section 2 "Methods and Materials" gives an overview of the research methodology and the experimental design, specifically, models, data, and software setup. Section 3 "Theory and Background" presents the relevant background in cybersecurity, industry 4.0/5.0, and industrial information integration. Section 4 "Numerical Results" presents the findings of the experiments. Section 5 "Discussion" analysis the findings in the context of the primary research question, which is how to integrate AI-enabled security solutions into a modern digital ecosystem. The discussion provides an overview of the challenges, presents a solution, and motivates future research directions. The last section gives a summary and conclusion.

## 2 Methods and Materials

### 2.1 Research Methodology

The primary goal of the paper is to secure modern complex digital ecosystems with AI-enabled cybersecurity models, with a specific focus on how to overcome the integration challenges of those models. To answer this research question, a mix of qualitative and quantitative research was chosen as the most suitable approach, utilizing the following research methods:

- **Literature review:** The goal of the literature review was to identify the current status quo of AI/ML in cybersecurity, including a general understanding of application domains. The literature review considered the following materials/sources: (1) Research papers from relevant scientific journals, (2) industry research and white papers with a focus on world-leading consulting companies, and (3) official announcements and agendas of government entities. The main search terms – and slight variations thereof – used during the literature search were the following: Cybersecurity, Artificial Intelligence, Deep Learning, Machine Learning, Intrusion

- **Experiment:** The three different scenarios network security, mobile security, and IoT security are utilized to carry out a quantitative study to test the accuracy and resilience of AI-enabled models, which will later serve as input for the integration discussion. The goal of the experiment is to train the base classifiers logistic regression, random forest, gradient boosting, and deep learning on three different publicly available datasets within the cybersecurity domain and combine them via stacking to a model referred to as super learning. The idea of the quantitative experimental study is to produce generalizable knowledge about the application of machine learning in cybersecurity. The free availability of the datasets allows reproducibility of the results for future research attempts in this area. This will assure research integrity and allow others to have trust and confidence in the methods used.
- **Conceptual analysis:** The findings of the experimental study will serve as crucial input for the analysis of the integration challenges and possible solutions. IIIE, which is defined as "a set of foundational concepts and techniques that facilitate the industrial information integration process", and "comprises methods for solving complex problems when developing IT infrastructure for industrial sectors, especially with respect to information integration" [46], will build the foundation to derive practical solutions and motivate further research.

Ultimately, the discussion combines the findings of the literature, the experimental results, and the theoretical and conceptual foundation of III to suggest practical solutions to the integration issue and motivate future research in this area.

## 2.2 Machine Learning Models

The experimental study uses four different ML models: Logistic Regression (LR), Random Forest (RF), Gradient Boosting Machine (GBM), and Deep Learning (DL).

(1) The Logistic Regression (LR) belongs to the family of generalized linear models (GLMs) and is widely used in the context of predictive analytics and binary classification in academia and industry.
(2) The recursive partitioning algorithm random forest creates different decision trees and averages the results in the end to reduce the variance of the prediction model. It is one of the most potent ML algorithms for classification and regression tasks out there.
(3) Gradient boosting in contrast does not build different trees and averages the outcomes but operates in sequential order. The specific idea of boosting is to start with a so-called weak learner – a model only slightly better than random guessing – that gradually improves by correcting the error of the previous model at each step. The most common form of boosting uses decision trees and sequentially ads one tree at a time. This step-by-step adjustment forces the model to gradually improve its performance and leads to higher accuracy [47]. There are several different gradient boosting implementations out there. This study uses the gradient boosting version implemented by Malohlava and Candel [48], which is based on Hastie et al., [47]. GBM can be considered state-of-the-art when it comes to prediction accuracy for supervised learning problems on structured data sets [49].

(4) Advances in AI research have improved the capabilities of artificial neural networks, which have started a new era of deep Learning [50]. The three factors mainly responsible for DL to become mainstream are increased data availability (Big Data), processing power (GPUs), and better optimization algorithms [51]. One of the major advantages of DL is its ability to work with unstructured data sets, which had a major impact in terms of breakthroughs in text, speech, image, video, and audio processing. Deep Learning can be used with different architectures such as feed-forward artificial neural networks (ANN), Convolutional neural networks (CNNs), as well as Recurrent Neural Networks (RNNs). However, the best architecture for tabular data – which is the case for our datasets – is a multi-layer feedforward artificial neural network. A deep learning model can consist of several hidden layers and is trained with stochastic gradient descent and backpropagation [51]. Applying a non-linearity in the form of an activation function is essential for neural networks to be able to learn complex (non-linear) representations of the input data sets. The activation function used for the hidden layers for the experiments in this paper is the rectified linear unit (ReLU).

(5) Super learning requires several pre-trained models, which can be combined via a fusion process into a more powerful super learner. Stacking, also known as super learning, is an ensemble method that combines several base classifiers into a more powerful super earner. In contrast to boosting, the goal of stacking is not a gradual improvement over weak learners. It takes instead several completely trained classifiers and merges them into a single stronger learner to increase prediction accuracy [28]. The idea of stacking was originally introduced by Wolpert [52] and later formalized by Breiman [53]. The theoretical underpinning was developed by Van Der Laan et al., [54], which proved that the model created through stacking – the super learner ensemble – represents an asymptotically optimal system for learning. In the first step, the base classifiers must be trained. Once completed, those predictions are fed into the meta learner to generate the final ensemble prediction [28].

## 2.3 Data and Preprocessing

This paper uses three different datasets to create a diverse range of application scenarios: network intrusion detection, malware detection, and intrusion detection within an IoT ecosystem. The datasets used in this study contain a wide variety of different attack types - see section 3.

**Table 1.** Description of Datasets

| Use Case | Observations | | | | | Description |
|---|---|---|---|---|---|---|
| | Total | y = 0 | y = 1 | Balanced | Features | |
| Network Intrusion Detection | 25,192 | 13,449 | 11,743 | 11743/11743 | 41 | Prediction of whether network traffic is usual behaviour or should be categorised as an attack |
| Android Malware Detection | 15,036 | 9,476 | 5,560 | 5560/5560 | 215 | Prediction whether an android application is malicious or benign |
| IoT Cyber Threat Detection | 157,800 | 24,301 | 133,499 | 24301/24301 | 61 | Prediction whether an action within an IoT ecosystem/on IoT device is malicious or benign |

*For the purpose of this study random under-sampling was used to bring the datasets in a balanced state

**Dataset 1 – Network Intrusion Detection:** The first dataset contains a total of 25192 observations, where 11743 are flagged as attack and 13449 as normal traffic. Each observation contains 41 features referring to the traffic input across a diver's range of categories including a response column that indicates whether the observation is normal traffic behavior or an attack.[1] The NSL-KDD dataset contains the following four attack types [55]: Denial of Service (DoS), Probe, User to Root(U2R), and Remote to Local (R2L). The dataset in its full form encompasses 125973 observations. For this study was the reduced version used, which is a 20% spin of the original dataset.[2]

**Dataset 2 – Android Malware Detection:** The second dataset (Drebin-215) represents observations from 15036 android applications, of which 9476 are benign apps (good cases) and 5560 are flagged as malicious. Each observation (or feature set) contains 215 extracted attributes/features including a binary response column, which can be either benign or malicious.[3] It was used in the paper "DroidFusion: A Novel Multilevel Classifier Fusion Approach for Android Malware Detection" to develop a multilevel classifier fusion approach for mobile malware detection [56].

**Dataset 3 – IoT Cyber Threat Detection:** The third and last dataset contains a total of 157800 observations, where 133499 are flagged as malicious and 24301 as benign. Each observation contains

---

[1] The NSL-KDD dataset for intrusion detection can be accessed here: https://www.unb.ca/cic/datasets/nsl.html
[2] A deep dive in the NSL-KDD dataset can be found here: https://towardsdatascience.com/a-deeper-dive-into-the-nsl-kdd-data-set-15c753364657
[3] The android malware detection dataset (Drebin-215) can be accessed here: https://www.kaggle.com/datasets/shashwatwork/android-malware-dataset-for-machine-learning

61 features including a response column that indicates whether the observation is normal behavior or an attack. The Edge-IIoTset dataset [57] contains a vast number of generated observations of different attack categories: DoS/DDoS attacks, Information gathering, Man in the middle attacks, Injection attacks, and Malware attacks. It is a modern and comprehensive dataset for research in industrial IoT security. The data are created from various IoT devices such as digital sensors for sensing temperature and humidity, ultrasonic sensors, water level sensors, pH sensor meters, heart rate sensors, and flame sensors [57]. For this study, the smallest ML dataset was used, which is still larger than the other two datasets.[4]

## 2.4 Software Setup

Data preparation and handling are managed in RStudio, which is the integrated development environment (IDE) for the statistical programming language R. R is one of the go-to languages for Data Science research as well as prototyping in practice. The machine learning models in this paper utilize H2O, an open-source platform written in Java that supports numerous predictive models [58]. The high abstraction level of this package emphasizes the idea and data, simplifying experimentation and streamlining problem-solving.

# 3 Theory and Background

## 3.1 Technologies and Security

### 3.1.1 Network Security

Networks are one of the basic building blocks of our modern digital and hyperconnected world. The omnipresence of global communication networks enables us to connect our devices, which allows us to get unprecedented access to a vast number of applications and services – globally and in real time. The idea of a centralized network perimeter gradually disappears due to a shift towards location independence and an IT infrastructure that is largely outsourced to a cloud environment [4,59]. Nevertheless, networks, whether wired or wireless are the primary enablers of our modern hyperconnected world. Information exchange between two devices over a network is done by sending packets back and forth, which is an attractive target for cybercriminals. Examples of attacks are packet sniffing, man-in-the-middle attacks, or distributed denial of service attacks, which try to compromise confidentiality, integrity, and availability. Network security measures are firewalls and virtual private networks (VPNs). Firewalls limit access to networks or devices by using access policies and rules that allow communication only with certain other networks or devices. A VPN prevents active and passive attacks by establishing an encrypted connection (tunnel) over the internet – it is the best network security measure available.

---

[4] The Edge-IIoTset dataset can be accessed here: https://ieee-dataport.org/documents/edge-iiotset-new-comprehensive-realistic-cyber-security-dataset-iot-and-iiot-applications

### 3.1.2 Mobile Security

Mobile devices such as smartphones and laptops have become an omnipresent reality in our modern world [60]. The economic shock suffered in 2020 caused by the coronavirus pandemic has further accelerated remote work. The extent of mobile devices used to access corporate systems is growing at a rapid pace and will with a high likelihood gain momentum over the years [61].

At the same time, the attack surface for smartphones and other mobile devices has grown and made those devices more vulnerable than ever to cyber threats [62]. Mobile devices are subject to all modern cyber threats such as malware, phishing, network attacks, supply chain attacks, and attacks on passwords and authentication.

Depending on the definition of a mobile device, we could consider any device that does not necessarily have a fixed physical location and can connect to the internet and is hence closely related to IoT devices. A distinction is not necessary, but would also be difficult. Smartphones are often the central communication hub and serve as a remote control for all kinds of smart devices in an IoT ecosystem.

### 3.1.3 IoT/CPS Security

The endless increase in smart devices has started seamlessly bridging the physical and digital worlds. The internet of things (IoT) is a broad concept and captures the idea of internet-enabled devices that seamlessly connect. It is a fundamental concept that enables cyber-physical systems, digital twins, and the (industrial) metaverse. Since the definition is very broad mobile devices and laptops are also IoT devices but are - for obvious reasons - part of mobile system security. Devices that usually count when we talk about IoT are all kinds of smart devices that enable smart homes, factories, and cities (smart TVs, smartwatches, cars, refrigerators, medical devices, and sensors - that transfer information from the real to the digital world, etc.). Essentially all devices that can establish a network connection to the internet or other devices.

The intersecting nature of IoT systems, including the numerous components required in the deployment of such systems comes with added security challenges [34,63]. IoT security is essentially a combination of network security, wireless security, and mobile system security [34]. IoT devices tend to have weak security software – especially compared to more mature devices such as smartphones and laptops – and offer adversaries a new attack vector. Smartphones are often used as a remote control for all kinds of smart devices. Cyber attackers that can gain control over the phone via a Remote Access Trojan (RAT) could control all IoT devices connected to the network as they are all already authenticated. Also, it is possible to use IoT devices as an easy access point into the network and carry out a lateral attack to infect other devices such as mobile phones, laptops, or other systems within the network and gradually gain full access to the target's activities, information, and data.

Cyber-physical systems are a closely related concept to IoT and describe systems that seamlessly integrate digital capabilities and physical devices and/or systems. The logic of smart cities and digital factories/industries is the same as IoT. Several smart devices communicate with each other, and an infection can easily spread through the network of the infected devices. IoT devices and interconnectivity have become complex and significantly increased the attack surface for cybercriminals. The strong integration between the digital and physical world allows cybercriminals to leverage network and device access to potentially carry out harmful actions in the real world. Attacks on critical infrastructure and cyber-physical systems (e.g., Triton, Stuxnet) are on the rise and can lead to significant damage.

### 3.1.4 Attacks & Malware

This part gives an overview of the attack types that exist in cyberspace with a focus on the categories that are contained in the used datasets. The IoT dataset uses a comprehensive array of attacks present in an IoT ecosystem and spans distributed denial of service (DDoS) attacks, information gathering, man-in-the-middle (MITM) attacks, injection attacks, and malware attacks – see Table 2 [57].

**Table 2.** Attack types present in the IoT dataset

| Category | Attack |
| --- | --- |
| DDoS attacks | TCP, UDP, HTTP, or ICMP flood |
| Information gathering | port scanning, OS fingerprinting, vulnerability scanning |
| MITM attacks | DNS spoofing, ARP spoofing |
| Injection attacks | cross-site scripting (XSS), SQL injection, uploading attack |
| Malware | backdoor, password cracking, ransomware attack |

DDoS attacks flood the system with an abnormal amount of traffic that ultimately causes a system to shut down for its own protection. This attack is quite common and uses botnets to increase attack strength and hide the source.

People or devices exchange information with another party via a network. A man-in-the-middle attack (MITM) secretly intercepts this communication channel by redirecting the traffic. This allows the adversary to take an active role, instead of only passively eavesdropping. This is often achieved via rogue Wi-Fi access points and evil twins. If the victim establishes a connection to this malicious access point the adversary could eavesdrop or modify the network communication [64].

Malware is one of the key components used in most criminal activities in cyberspace and it continues to grow in volume and complexity posing significant threats to the security of corporations, governments, and individuals. With its help, attackers can achieve the full range of confidentiality, integrity, and availability violations. It is often categorized as a separate attack or risk type; however, malware is part of most cyber-attacks. Sometimes as an enabler and sometimes as a package to be delivered. It has a very broad spectrum of utilization. (1) DoS attacks use botnets (malware); (2) Phishing is often used to place malware on the host devices to carry out several different types of malicious actions; (3) Ransomware attacks utilize sophisticated malware packages to carry out the operation.

The primary goal of malware is to infiltrate, disrupt, and/or damage computers and systems without the end-user's approval. It can manifest itself in every layer of the system stack as an independent program or embedded in a host application [65]. There a several different malware types including viruses, worms, trojans, backdoors, rootkits, spyware, adware, keyloggers, botnets, and more.

## 3.2 Digital Ecosystems

### 3.2.1 Industry 4.0 and Industry 5.0

Humankind has witnessed many industrial transformations [5,7]:

(1) the first industrial revolution – the steam age,
(2) the second industrial revolution – the age of electricity,
(3) the third industrial revolution – the information age,
(4) the fourth industrial revolution – the age of cyber-physical systems, and
(5) the fifth industrial revolution – mass personalization.

The primary current and emerging enabling technologies of both – industry 4.0 and industry 5.0 – are IoT, CPS, AI, cloud/edge computing, modern mobile networks such as 5G/6G, blockchain, and quantum computing [7,66]. The idea of industry 5.0 is an extension of and hence includes industry 4.0 concepts with a stronger focus on circularity and human empowerment in an increasingly digital world economy. Overall, enabling the seamless integration and interoperability of all those technologies and concepts on a global scale are at the heart of digital transformation and the transformation to industry 4.0 and 5.0. The essence of industry 4.0 and industry 5.0 is the integration of those technologies into a digital enterprise – or more generally – a complex digital ecosystem.

### 3.2.2 Industrial Information Integration

Industrial Information Integration Engineering (IIIE) is an emerging and relatively new discipline. Xu [46] frames the conceptual, theoretical, technological, and engineering foundations of IIIE along with its interdisciplinary structure [46]. It focuses on how to integrate and manage various systems and data in an industrial context and combines concepts and methods from systems engineering, data science, computer science, and industrial engineering to create efficient, effective, and robust industrial information systems. The foundational building blocks of IIIE are [5,46]:

- **Enterprise Architecture (EA)**: EA helps to align business objectives with IT strategy, processes, and infrastructure. It serves as a blueprint and provides a comprehensive view of the interrelationships of an organization's information systems and technology, enabling more effective decision-making and technology investment.
- **Enterprise Application Integration (EAI)**: EAI is a framework for integrating a set of enterprise applications, enabling data flow across disparate applications within the enterprise. It helps to reduce data silos, enhances interoperability, and facilitates streamlined business processes through process and information integration.
- **Service Oriented Architecture (SOA)**: SOA is an architectural style that structures an application as a collection of services that are easy to integrate and reuse. It is an architecture for integrating platforms, protocols, and legacy systems and is a primary enabler of heterogeneous systems as it enhances flexibility, promotes interoperability, and accelerates business processes.
- **Business Process Management (BPM)**: BPM is a systematic approach to optimize and/or automate an organization's workflow to make it more efficient, effective, and flexible.
- **Information Integration and Interoperability**: Integration has become a core competence of digital transformation. In industry 4.0 there are different integration types as vertical integration, horizontal integration, and end-to-end digital integration, which aim to reduce data silos and ensure that information can be effectively shared and utilized across various systems and applications.

## 4 Numerical Results

This section presents the numerical results of the three threat detection scenarios: network intrusion, android malware, and IoT. The three scenarios were carefully chosen to cover the primary building blocks of our digital world. Network traffic - whether wired or wireless - is one of the crucial underpinnings of our globally interconnected world. Mobile devices offer constant connectivity and availability of individuals, applications, and systems and have fundamentally changed the way we live and work. IoT and CPS are broad concepts that capture the idea of digital ecosystems and enable the fusion of the digital and physical worlds.

The experiments compare the performance of the two proposed super learners against the four base classifiers logistic regression, random forest, gradient boosting, and deep learning. The evaluation metrics used for the comparisons are AUC, accuracy, and F-score. The accuracy and F-scores are reported at a 0.5 threshold level. The strongest base classifier and the strongest super learner in terms of the evaluation metrics are highlighted in black. The logistic regression classifier was not used during the training of the super learners SL1 and SL2.

## 4.1 Scenario 1: Network Intrusion Detection

Networks, whether wired or wireless are the primary enabler of our modern hyperconnected world. The numerical results for dataset 1 - network intrusion detection - predict whether network traffic is usual behavior or should be categorized as an attack. Table 3 shows clearly that gradient boosting has the best overall performance with the highest AUC, accuracy, and F-score of 1.000, 0.9966, and 0.9966 respectively.

**Table 3.** Numerical results for Dataset 1 - Network Intrusion Detection

| Classifier | Candidate | Out-of-Sample Performance | | |
|---|---|---|---|---|
| | | AUC | Accuracy | F-score |
| Logistig Regression | | 0.9900 | 0.9454 | 0.9451 |
| Random Forest | | 0.9999 | 0.9945 | 0.9944 |
| **Gradient Boosting** | | 1.0000 | 0.9966 | 0.9966 |
| Deep Learning | | 0.9996 | 0.9860 | 0.9859 |
| SL1: DL | RF, DL, GBM | 1.0000 | 0.9966 | 0.9966 |
| **SL2: GBM** | RF, DL, GBM | 1.0000 | 0.9970 | 0.9970 |

In terms of AUC maximum performance has been already achieved during the base classifier training. The random forest comes as a close second with an AUC of 0.9999. Both ensemble models achieve a better performance in the case of the network intrusion dataset than the DL model with an AUC of 0.9996 and also lower accuracy and F-score.

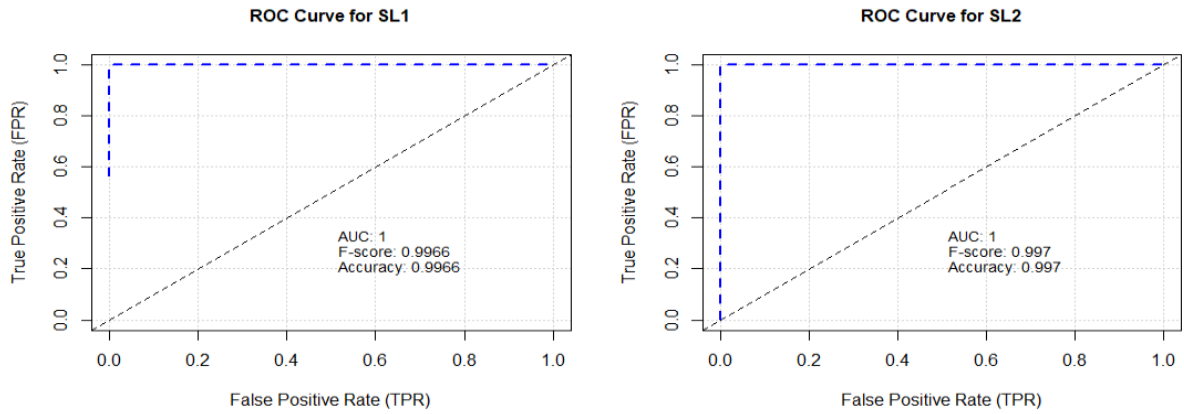

**Figure 2.** Scenario 1 – Network Intrusion Detection - ROC curve and evaluation metrics for SL1 (deep learning) and SL2 (gradient boosting). Network intrusion detection is largely concerned with identifying malicious network traffic to prevent cybercriminals from gaining access to internal devices or resources. The goal was to predict whether network traffic is usual behavior or should be categorized as an attack.

Figure 2 shows the ROC curve for SL1 and SL2 including the evaluation metrics for the network intrusion dataset. Both achieve a perfect AUC, which is no improvement compared to the single GBM classifier. However, the accuracy and F-score for the GBM super learner are slightly higher.

### 4.2 Scenario 2: Android Malware Detection

Malware is one of the key attack vectors that is consistently utilized to compromise systems and is part of most cyber attacks. The numerical results for dataset 1 - android malware detection - presented in Table 4 show clearly that gradient boosting has the best overall performance with the highest AUC, Accuracy, and F-score of 1.000, 0.9966, and 0.9966 respectively.

**Table 4.** Numerical results for Dataset 2 - Android Malware Detection

| Classifier | Candidate | Out-of-Sample Performance | | |
|---|---|---|---|---|
| | | AUC | Accuracy | F-score |
| Logistig Regression | | 0.9930 | 0.9714 | 0.9706 |
| Random Forest | | 0.9968 | 0.9794 | 0.9787 |
| **Gradient Boosting** | | 0.9981 | 0.9794 | 0.9788 |
| Deep Learning | | 0.9979 | 0.9767 | 0.9763 |
| **SL1: DL** | RF, DL, GBM | 0.9985 | 0.9821 | 0.9816 |
| SL2: GBM | RF, DL, GBM | 0.9984 | 0.9812 | 0.9807 |

The two best base classifiers in this scenario are gradient boosting and deep learning. The random forest comes and takes third place.

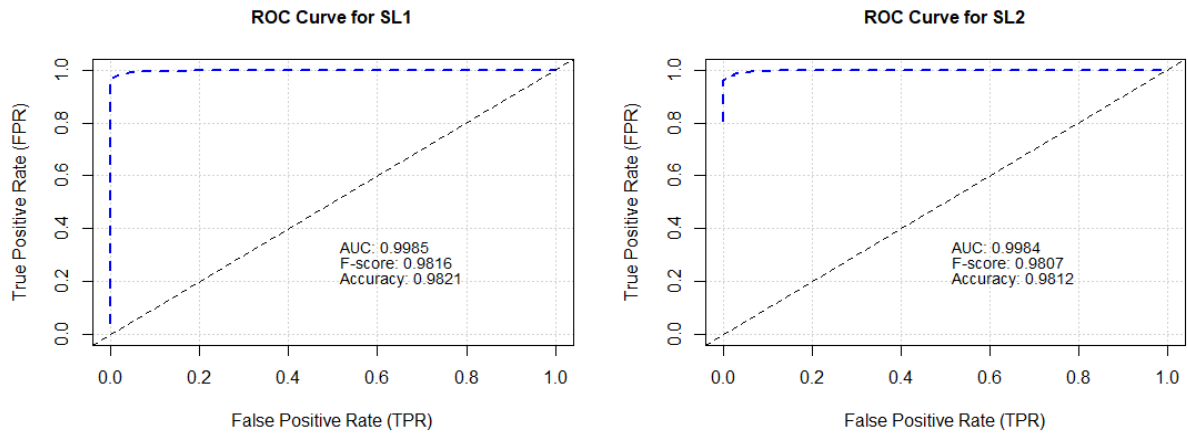

**Figure 3.** Scenario 2 - Android Malware Detection - ROC curve and evaluation metrics for SL1 (deep learning) and SL2 (gradient boosting). Malware is designed to infiltrate, disrupt and/or damage computers without the user's consent and is one of the key components used in most criminal activities in cyberspace. The goal was to predict whether an android application is malicious or benign.

Figure 3 shows the ROC curve for SL1 and SL2 including the evaluation metrics for the android malware dataset. Both outperform the base classifiers in terms of AUC, accuracy, and F-score. This is the only scenario where the DL super learner achieves a higher performance across all three evaluation measures compared to the super learner based on gradient boosting.

## 4.3   Scenario 3: IoT Cyber Threat Detection

IoT is a broad concept and captures the idea of internet-enabled devices that seamlessly connect to the digital and physical worlds. It is a fundamental concept that enables cyber-physical systems, digital twins, and the (industrial) metaverse. IoT devices tend to have weak security software and offer adversaries a new attack vector. The numerical results for dataset 3 – IoT cyber threat detection presented in Table 5 are similar to the network intrusion detection dataset.

**Table 5.** Numerical results for Dataset 3 - IoT Cyber Threat Detection

| Classifier | Candidate | Out-of-Sample Performance | | |
| --- | --- | --- | --- | --- |
| | | AUC | Accuracy | F-score |
| Logistig Regression | | 0.8747 | 0.7945 | 0.7817 |
| Random Forest | | 0.9975 | 0.9876 | 0.9878 |
| **Gradient Boosting** | | 0.9990 | 0.9856 | 0.9859 |
| Deep Learning | | 0.9879 | 0.9433 | 0.9437 |
| SL1: DL | RF, DL, GBM | 0.9995 | 0.9897 | 0.9899 |
| **SL2: GBM** | RF, DL, GBM | 0.9996 | 0.9903 | 0.9905 |

GBM is the winner in terms of performance with the highest AUC, accuracy, and F-score of 0.9990, 0.9856, and 0.9859 respectively. RF takes the second place with an AUC, accuracy, and F-score of 0.9975, 09876, and 0.9878. Both ensemble models achieve better performance for the IoT dataset than the DL model.

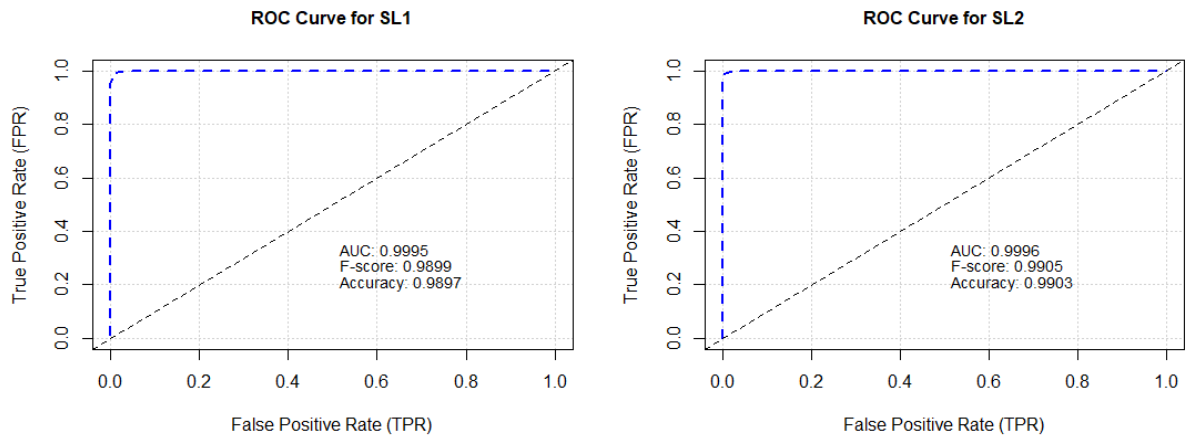

**Figure 4.** Scenario 3 - IoT Cyber Threat Detection - ROC curve and evaluation metrics for SL1 (deep learning) and SL2 (gradient boosting). IoT is a broad concept and captures the idea of internet-enabled devices that seamlessly connect to the digital and physical worlds. IoT devices tend to have weak security software and offer adversaries a new attack vector. The goal was to predict whether an action within an IoT ecosystem/on an IoT device is malicious or benign.

Figure 4 shows the ROC curve for SL1 and SL2 including the evaluation metrics for the IoT dataset. Both outperform the base classifiers in terms of AUC, accuracy, and F-score. The super learner based on gradient boosting achieves a higher performance across all three evaluation measures compared to the super learner based on deep learning.

## 5 Discussion

The modern technology-driven world requires intelligent security systems to protect our diverse digital industries and smart infrastructures. The experiment in this paper analyzed AI/ML-based intrusion detection on three datasets covering the scenarios of network security, mobile security, and IoT/CPS security. While choosing the right model based on resilience, and robustness are important for capable security solutions, integrating those models into complex digital ecosystems often poses a bigger challenge. Industrial Information Integration Engineering (IIIE) could be extremely helpful when deploying or integrating AI-enabled cybersecurity solutions into existing enterprise systems and IT infrastructures.

## 5.1   Model choice, prediction accuracy, and resilience

*RQ1: What is the best AI/ML model for anomaly-based supervised intrusion detection systems to detect cyber-attacks on mobile and IoT-enabled smart devices?*

The answer to this question was already motivated during the literature review. Based on several recent reviews [29,36,67] in the field of AI-enabled IDS and according to the results presented in this research paper, we can conclude that a clear go-to AI/ML model in terms of a clear winner does not exist. Based on the tested metrics AUC, accuracy, and F-score, we can see that gradient boosting could achieve the highest accuracy out of the tested models for the network security, android malware, and IoT security scenarios. However, RF and DL also offered strong performance in terms of those metrics. In general, the classification accuracy delta between DL and ensemble models as GB and RF is not necessarily significant and both seem to be able to offer strong performance in different scenarios (e.g. DL was the best meta-learning for the android dataset). Those findings are in line with the current literature that seems to be consistent in the sense that there is no clear go-to model. Shukla et al., [42] also mention that attack detection and mitigation are fundamental parts of model resilience, but their effectiveness is greatly dependent on the assumed context and attack models, which is in line with the findings in this article.

In addition, the experiments based on the three application scenarios network intrusion detection, android malware detection, and IoT cyber threat detection demonstrated that we can further enhance and optimize prediction accuracy for ML-classifiers within cybersecurity by utilizing a fusion strategy as stacking ensembles to create a super learning model. The numerical results presented are across all three datasets in favor of gradient boosting as the base classifier. When it comes to the super learner gradient boosting also turns out to be the winner except for the android malware dataset where deep learning was the strongest meta-learner. The super learners were trained by utilizing the base classifiers random forest, gradient boosting, and deep learning. It turns out that adding "weak" classifiers to the overall mix of the solutions dilutes the prediction accuracy of the final super-learner, hence the logistic regression was dropped as it had the weakest performance and did not add additional value.

## 5.2   Integration challenges of AI-enabled security solutions

*RQ2: What are the primary challenges when integrating AI-enabled cybersecurity models in complex enterprise ecosystems?*

Integrating AI/ML applications for cybersecurity in a digital ecosystem at the enterprise level is a complex undertaking. Table 6 presents challenges across the dimensions of process, data, application, and infrastructure. The concepts and techniques of industrial information integration could help to effectively integrate AI-enabled cybersecurity models into complex enterprise systems. Leveraging IIIE principles and overcoming the challenges presented in Table 6 will facilitate the seamless integration and adoption of advanced cyber threat detection solutions, ultimately enhancing the security of critical systems and networks and hence the robustness and resilience of the overall digital ecosystem.

**Table 6.** Integration challenges of AI-enabled cyber threat detection in complex digital ecosystems

| Dimension | Integration challenges |
|---|---|
| Process | - **Complexity of ML models**: ML models for threat detection can be quite complex, requiring a high level of expertise to develop, integrate, and maintain.<br>- **Continuous learning and adaptation**: Cyber threats evolve rapidly, so ML models must be continually updated and trained with new data to stay effective.<br>- **Evaluation of effectiveness**: Determining effectiveness in terms of false positives and negatives. |
| Data | - **Quality and availability**: ML models require large volumes of high-quality data for training. Obtaining this data can be challenging, especially in host-based IDS where data privacy concerns are significant.<br>- **Imbalanced data**: Cybersecurity datasets are often imbalanced, with many more examples of normal behavior than of attacks. This can make it difficult for ML models to learn to detect threats effectively.<br>- **Data privacy and security**: Data used for training and validation must be carefully managed to ensure privacy and security. |
| Application | - **Interoperability:** ML models need to be able to interact effectively with various applications and systems in an enterprise. This can be particularly challenging in network-based IDS where the diversity of network traffic and protocols is high.<br>- **Real-time threat detection:** Cyber threat detection needs to be performed in real-time, which can be difficult for ML models that require significant computational resources. |
| Infrastructure | - **Scalability:** The infrastructure needs to be robust and scalable to handle the high data volumes and computational demands of ML models.<br>- **Deployment environments:** Host-based IDS need to operate effectively on a wide range of system configurations, while network-based IDS need to handle diverse network architectures and protocols.<br>- **Resource constraints:** ML models can be resource-intensive, which can pose challenges for deployment, particularly in host-based IDS where system resources may be limited (e.g. IoT devices) |

### 5.2.1 Network security

Network security is the practice of protecting an organization's network infrastructure from various types of cyber threats. The integration of AI-enabled solutions here is challenged by the dynamic and complex nature of modern network environments, where traffic patterns are continuously evolving and the line between internal and external networks is often blurred due to cloud adoption and remote work. We are living in a cloud-first world and more and more applications will migrate from an on-premises to a cloud environment. Successful integration needs to ensure that AI models can effectively analyze vast amounts of network data in real-time, adapt to changing patterns, and respond rapidly to detect threats. This requires sophisticated data integration and processing capabilities, as well as robust AI algorithms that can learn from evolving network behavior. The unique challenge here is dealing with the interconnectedness of network components, where a single point of weakness can compromise the entire system. However, the movement towards a cloud-first world plus the dislocation of the workforce from the office (internal network) to remote working places (outside the network) makes it vital to rethink how to approach network security in the future. The different stakeholders can now directly communicate with the cloud without first connecting to the central network and zero-trust security solutions are vital to secure our decentral digital ecosystems [68,69].

### 5.2.2 Mobile security

The old world of network perimeters is gradually evaporating. The world has become decentral, applications have moved to the cloud, and employees are mobile. Hence, mobile devices have become the primary access point to the enterprise system, and connecting to a central network is often not necessary anymore. The internet is the new corporate network where cloud solutions act as a new gravitational force. In this decentral world, a concept referred to as zero trust has become a standard approach to security in many organizations, especially as remote and hybrid work become more prevalent. Zero trust enables granular user-access control through network segmentation and prevents lateral movements. It centered on the belief that organizations should not automatically trust anything inside or outside their perimeters. This means that all devices, users, apps, and network traffic – whether internal or external – are considered potentially threatening until verified [68,69].

The distributed nature of those mobile devices has increased the integration complexities. First, it increases the need for host-based security solutions. The primary challenge here is the sheer diversity of mobile device types, operating systems, and applications, along with the need to balance security with user experience. Second, AI-enabled solutions must be capable of detecting threats in a decentralized and heterogeneous environment, often without the continuous connectivity that is typically available in network security. Third, privacy concerns are acute in this scenario, as AI systems need to monitor device usage patterns without violating privacy.

### 5.2.3 IoT/CPS security

The unique characteristic of IoT/CPS security is the convergence of the digital and physical worlds, necessitating cybersecurity solutions that can ensure both data integrity and physical safety. Also, the logic of an increasingly decentralized and the need for host-based security solutions is similarly true for IoT/CPS security. Many IoT devices have significant resource limitations, which makes it difficult to deploy power-hungry DL models. Minimal model complexity is highly important for devices that often have weak CPUs. In the literature review, it was shown that many different AI/ML models deliver adequate attack prediction accuracy to be a viable choice for the final deployment on those devices. Shukla et al., [42] for example, have shown that neural networks with a single layer are performing well already. Thus, neural networks are viable for use in attack detection scenarios within an IoT environment. Since a neural network with a single hidden layer does deliver adequate attack prediction accuracy within IoT environments, we can infer that most other ML classifiers, e.g., random forest, and gradient boosting, should be able to deliver similar strong results. Indeed, the experiments in this paper have demonstrated that DL is not necessarily the best solution, and faster models, which require less CPU power are not only on par with DL but have the potential to perform better. Overall, the main integration challenge in this scenario is the vast number and diversity of devices, many of which have limited computing power and may not have been designed with security as a priority. AI-enabled solutions must be lightweight enough to run on resource-constrained devices, yet sophisticated enough to detect complex threats. Data privacy and reliability are also significant concerns, especially given the real-world impact of potential security breaches in CPS (e.g., healthcare).

## 5.3 Towards integrated, secure, and resilient digital ecosystems

### 5.3.1 Utilizing ML pipelines to integrate AI-enabled security solutions

*RQ3: How can we successfully integrate AI/ML-based cyber threat detection systems into complex digital ecosystems?*

In addressing the challenges of integrating AI-enabled cybersecurity models into complex enterprise systems, the concepts and techniques of industrial information integration [46] are essential, especially when leveraged in conjunction with a well-structured machine-learning pipeline [28]. The incorporation of an ML pipeline allows for consistent and efficient data preprocessing, model training, evaluation, and deployment. Such a pipeline setup will help to ensure data quality and availability across the enterprise, which is vital for model performance. By improving interoperability between disparate systems, IIIE, and the machine learning pipeline together allow for seamless integration and interaction of AI models with various elements of the enterprise system. They can facilitate scalability by ensuring that the data infrastructure and hence the ML models can handle the high volumes of data that exist in modern distributed enterprise ecosystems. This can be achieved through cloud-based solutions and distributed processing. The integration of this ML pipeline and the AI-enabled security solutions with the overall enterprise IT architecture also plays a critical role in enabling real-time data analysis, which is crucial for prompt threat response. The ML pipeline also helps to reduce false positives and negatives by ensuring data quality due to rigorous validation processes, and it could address potential bias in AI models through representative data collection and bias detection and correction mechanisms. However, this part is rather tricky since capable adversaries will invent new attacks (e.g., zero days) or use other concealment techniques to circumvent security solutions. Despite the sophisticated nature of many adversarial attacks, the combination of IIIE and a well-defined ML pipeline could potentially enhance the system's overall robustness by establishing more robust security protocols and techniques for detecting such threats. It has been shown that AI-enabled cybersecurity certain extent capable of identifying zero-day attacks. This integrated setup would also ensure easy system maintenance and updates through modular designs, containerization, and regular system checks. Overall, the principles of IIIE together with an ML pipeline could positively impact the resilience, robustness, and potentially explainability of AI-base security systems.

### 5.3.2 Limitations and Future Research

Future research to strengthen the robustness, response, and resilience of AI-based cyber security systems is vital to secure our modern digital ecosystems. One key issue with AI in cybersecurity is that there is no guarantee that the AI algorithm – due to obfuscation and concealment tactics from cybercriminals – will capture the threat. In addition, it might misclassify normal behavior as malicious, which could result in unnecessary expenses.

Future research should focus on improving the state-of-the-art threat and anomaly detection accuracy to increase the resilience of AI in cybersecurity. A specific focus on cyber-physical systems (CPS) and internet of things (IoT) devices is a worthwhile research area in general due to a significant increase in smart devices [11], infrastructures, and the move towards smart cities, digital twins, and the (industrial) metaverse. Cyber-physical systems security for advanced human adventures (e.g. space travel, colonization of Mars) will be vital during the coming decades [70].

Trust in AI in cybersecurity is vital to mitigating the double-edged sword problem. Many situations are unfortunately difficult to avoid. Social engineering (phishing) in particular, which relies on a failure in human behavior is a good example. Especially APTs tend to utilize advanced forms of whale phishing that are difficult to prevent. Since attacks of those types are usually long-term in nature, an initial failure will not result in the termination of the attack. The attacker will continue until the objective has been achieved. Research here could focus on hardening/strengthening AI-based security systems against adversarial attacks (robustness). It is possible to increase the scope by adding additional datasets to increase the validity and generalizability of the work presented and therefore strengthen the contribution. Since the major focus was on network security, mobile security, and security in an industrial IoT environment, there would be the option to choose additional datasets that increase the breadth of the study. Since everything is in the cloud these days the traditional network security perimeter is largely a thing of the past and a concept called zero-trust security is the new way of protecting our systems and enterprises. Hence, a focus on end-point security makes sense. Another option is to add additional machine learning methods in addition to deep learning and gradient boosting. Also, there are options to increase the variations of deep learning by changing the activation functions or types of networks. A further option is to utilize different fusion methods to increase the prediction strengths/accuracy of the models. And finally, integrating AI algorithms as presented in this paper into more comprehensive security solutions is required for real-world applications in enterprises and governments. Further investigations on how IIIE can be used to address the integration challenges of AI-enabled cyber threat detection models, to overcome potential obstacles in the process, data, application, and infrastructure dimensions, is a worthwhile future research direction.

It is also possible to move away from enterprises and governments and focus on home network protection. While most research and funding go towards the protection of large corporates and governments, general society, especially home networks has received little attention so far. As technological advancements continue to accelerate the likelihood of attacks on individuals and home networks will further increase. Also, due to the trend toward remote work, and the gig economy end-user security for individuals will become increasingly important. And finally, I would like to motivate research in AI for human empowerment through AI-driven co-creation, where AI and humans act as partners.

Overall, the above-presented research directions are vital to increasing the robustness, response, and resilience of AI systems. We need to strengthen the classification power of AI and ensure the deployment and integration of reliable AI for cybersecurity solutions to protect our complex digital ecosystems – and ultimately our future.

# 6  Conclusion

The world and humanity are increasingly more connected and technologically advanced, and the natural boundaries between the digital and physical worlds have gradually ceased to exist. Mobile devices and communication networks offer constant connectivity and availability of individuals, applications, and systems and have fundamentally changed the way we live and work. Modern paradigms such as the internet of things and cyber-physical systems capture the idea of digital ecosystems and enable the seamless integration of those two worlds – digital and physical. While those developments are fascinating and open fast new possibilities for communication and interaction globally, this increased interconnectedness of all devices significantly increased the attack surface from a cyber security perspective. The pervasiveness of digital technologies and the constantly shifting technological foundation have created a complex and powerful playground for cybercriminals. And the increasing usage of all kinds of distributed smart devices has led to a significant surge in demand for intelligent threat detection systems based on machine and deep learning. This paper explored AI-based intrusion and malware detection to protect our modern digital ecosystems from cyber-attacks. It was shown that we can enhance the prediction accuracy of machine learning applications within cybersecurity to increase the resilience of AI-based defense systems for network security, mobile security, and IoT security.

Cybersecurity is an essential key enabler for all major endeavors going forward and Industrial Information Integration Engineering (IIIE) could be helpful when deploying or integrating those AI-enabled cybersecurity solutions into existing enterprise systems and IT infrastructures. Examples are the further digitalization of businesses, the protection of critical infrastructures, and securing government institutions, and individual end-users. But also, explorations to the moon and beyond will require sophisticated security solutions. The research possibilities in this area are fascinating and endless and automated intelligent systems are the only option to shield our world against cyberattacks that are highly funded, organized, complex, and as well powered by artificial intelligence. Cyber security will remain a constant fight against smart and rational adversaries, which are consistently striving to improve upon the status quo, and we should do our best to be prepared.